\documentclass[traditabstract]{aa} 

\usepackage{graphicx}
\usepackage{natbib}
\usepackage{colortbl}
\usepackage{color}
\usepackage{array}
\usepackage{amsfonts,amssymb,amsmath}
\usepackage{multirow}
\usepackage{txfonts}
\usepackage{lscape}

\newcommand{\dtm}{$\mathcal{{DT\!\!M}}$}
\newcommand{\dtma}{\textit{dtm}}
\newcommand{\hi}{\mbox{H\,{\sc i}}} 
\newcommand{\hii}{\mbox{H\,{\sc ii}}}

\newcommand{\feii}{\mbox{Fe\,{\sc ii}}} 
\newcommand{\siii}{\mbox{Si\,{\sc ii}}}

\newcommand{\suii}{\mbox{S\,{\sc ii}}}

\newcommand{\znii}{\mbox{Zn\,{\sc ii}}}

\newcommand{\nh}{$N(H)_{\rm{X}}$}
\newcommand{\avnh}{$A_V/N(H)_{\rm{X}}$}

\newcommand{\nfed}{$N$(Fe)$_{\rm dust}$}
\def\lya{Lyman-$\alpha$}

\begin{document}

\title{Dust-to-metal ratios in damped Lyman-$\alpha$ absorbers:}
\subtitle{Fresh clues to the origins of dust and optical extinction towards $\gamma$-ray bursts}

   \author{A. De Cia \inst{1,}\inst{2}, C. Ledoux \inst{3}, S. Savaglio \inst{4}, P. Schady \inst{4} and P.~M. Vreeswijk \inst{1}}

\institute{ 
Department of Particle Physics and Astrophysics, Faculty of Physics, Weizmann Institute of Science, Rehovot 76100, Israel \\
\email{annalisa.de-cia@weizmann.ac.il}
\and
Centre for Astrophysics and Cosmology, Science Institute, University of Iceland, Dunhagi 5, 107 Reykjavik, Iceland 
\and
European Southern Observatory, Alonso de C\'ordova 3107, Casilla 19001, Santiago 19, Chile
\and
Max-Planck Institut f\"ur Extraterrestrische Physik, Giessenbachstra\ss{}e 1, 85748, Garching, Germany
}
   
   \date{Received Month dd, 2013; accepted Month dd, 2013}

  \abstract{Motivated by the anomalous dust-to-metal ratios derived in the literature for $\gamma$-ray burst (GRB) damped \lya{} absorbers (DLAs), we measure these ratios using the dust-depletion pattern observed in UV/optical afterglow spectra associated with the interstellar medium (ISM) at the GRB host-galaxy redshifts. Our sample consists of 20 GRB absorbers and a comparison sample of 72 QSO-DLAs with redshift $1.2<z<4.0$ and down to $Z=0.002\,Z_\odot$ metallicities. The dust-to-metal ratio in QSO- and GRB-DLAs increases both with metallicity and metal column density, spanning $\sim$10--110\% of the Galactic value and pointing to a non universal dust-to-metal ratio. The low values of dust-to-metal ratio suggest that low-metallicity systems have lower dust fractions than typical spiral galaxies and perhaps that the dust in these systems is produced inefficiently, i.e. by grain growth in the low-metallicity regime with negligible contribution from supernovae (SNe) and asymptotic giant branch (AGB) stars. On the other hand, some GRB- and QSO-DLAs show high dust-to-metal ratio values out to $z\sim4$, requiring rapid dust production, such as in SN ejecta, but also in AGB winds and via grain growth for the highest metallicity systems. GRB-DLAs overall follow the dust-to-metal-ratio properties of QSO-DLAs, GRBs probing up to larger column and volume densities. For comparison,  the dust-to-metal ratio that we derive for the SMC and LMC  are $\sim$82--100\% and $\sim$98\% of the Galactic value, respectively. The literature dust-to-metal ratio of the low-metallicity galaxy I Zw 18 ($<37$\%) is consistent with the distribution that we find. The dust extinction $A_V$ increases steeply with the column density of iron in dust, \nfed{}, calculated from relative metal abundances, confirming that dust extinction is mostly occurring in the host galaxy ISM. Most GRB-DLAs display $\log \mbox{\nfed{}} >14.7$, above which several QSO-DLAs reveal molecular hydrogen, making GRB-DLAs promising candidates for molecular detection and study.}


\titlerunning{Dust-to-metal ratios in DLAs}

\authorrunning{De Cia et~al.}

   \maketitle

\section{Introduction}

The interplay between the production/recycling of metals and the formation/evolution of cosmic dust is strongly connected with star formation \citep[e.g.,][]{Gall11,Cortese12} and plays a fundamental role in the chemical enrichment of the interstellar medium (ISM). One way of estimating the dust content along the line of sight is through the optical dust extinction $A_V$ \citep[e.g.,][]{Pei92}. The low-ionization gas content is typically traced by the neutral hydrogen $N$(\hi{}), while a possible probe of the metals is the soft X-ray absorption \nh{} \citep[e.g.][]{Morrison83,Wilms00}. $A_V/N$(\hi{}) can then be used to infer the dust-to-\textit{gas} ratio. However, this ratio does not provide a natural way to quantify the amount of dust in a galaxy's interstellar medium as it does not include the metals in the gas. On the other hand, the dust-to-metal ratio, i.e. the fraction of metals in dust, represents the ability of a system to produce or destroy dust grains. Besides, it can be derived regardless of the \hi{} constraint, which is often not available (see Sect \ref{sample}). We refer to \dtm{} to indicate the dust-to-metal ratios normalized by the Galactic value.

The dust-to-metal ratio in the Galaxy is $A_V$/\nh{}[Gal] $ \sim5.6\times 10^{-22}$ mag cm$^2$ \citep[][the latter two providing a 20\% lower value]{Predehl95,Guver09,Watson11}, in agreement with the measurements based on UV absorption lines \citep{Bohlin78}. The SMC and LMC have $A_V$/\nh{}[SMC]\footnote{This value is derived from the observed $E(B-V)/N({\rm H})$ from \citet{Martin89} and \citet{Draine03} and converting it to \nh{}/$A_V$ assuming $R_V=2.9,3.2$ \citep{Pei92,Draine03} and a mean [Zn/H] $=-0.64,-0.55$ for the SMC and LMC respectively \citep{Welty97,Welty99}.} $ \sim2.8\times 10^{-22}$ mag cm$^2$ and $A_V$/\nh{}[LMC]\footnotemark[1] $\sim5.3\times 10^{-22}$ mag cm$^2$, corresponding to $\sim50$\% and $\sim~95$\% of the Galactic value, respectively. Such measurements are typically difficult to carry out for distant galaxies. \citet{Dai09} studied the ratio in gravitationally lensed galaxies at $z<1$ by comparing the extinction curve and the X-ray absorption in the spectra of multiply-imaged quasars (QSOs) in the background. They find the dust-to-metal ratio to range from $\sim10$\% to a few times the Galactic value, but with large uncertainties. \citet{Brinchmann13} recently derived dust-to-metal ratios increasing with metallicity for super-solar metallicity galaxies based on observed CO absorption and the theoretical expectations of \citet{Charlot01}.

Long-duration ($>2$~s) GRBs can offer the unique possibility of directly probing the gas, dust, and metal content in their star-forming host galaxy, out to very high redshifts. Their afterglows are often bright enough to allow absorption-line spectroscopy that can reveal the \hi{} gas and metal column densities in the ISM \citep[e.g.,][]{Savaglio03,Vreeswijk04,Berger06,Prochaska07,Ledoux09,Fynbo09}. In addition, the well-understood power-law continuum \citep[e.g.,][]{Sari98,Piran99} facilitates the determination of the intrinsic $A_V$ and \nh{} \citep[e.g.,][]{Campana06,Kruehler11,Zafar11,Schady12}. The latter is measured for most \textit{Swift} bursts \citep{Evans09} and $A_V$  can be studied out to $z>6$ \citep{Schady10,Zafar11b}. In contrast, the same measurements are more difficult to make towards QSOs, due to the presence of broad emission lines contaminating a large fraction of their optical/ultraviolet continuum. The samples of GRB absorbers that are typically followed up for spectroscopic studies are biased against the most dusty systems that have been dimmed by large $A_V$'s or faint GRBs. A fraction of GRB host galaxies is massive and dusty \citep{ Kruehler11, Hunt11, Rossi12, Perley13}. Nevertheless, the current (biased) samples are still representative of the majority of the GRB host population, i.e., low-mass, low-metallicity and actively star-forming galaxies \citep[e.g.][]{Savaglio09}.

\citet{Zafar11} and \citet{Schady10} investigated the dust-to-metal ratios for a large sample of distant GRB-selected galaxies using \avnh{}. They found \avnh{} in GRBs that were up to two orders of magnitude lower than what had been observed in the Local Group, confirming the results of \citet{Galama01}. This implies that either GRB-DLAs have a different dust-to-metal ratios or that $A_V$ and \nh{} probe different regions along the line of sight. While the dust is expected to lie in the cold low-ionization medium, the X-ray absorption traces the metals regardless of their ionization state. The evidence of ionization induced by the GRB on the surrounding medium, in the immediate environment of the GRB for most bursts and out to hundreds of pc for absorbing systems with low $N$(\hi{}) \citep{DeCia12,Vreeswijk13}, confirms that indeed the line of sight analysis can be affected by ionization and dust destruction. Thus, the \avnh{} may not provide the best estimate of the dust-to-metal ratio for GRB host galaxies. Recent estimates based on line of sight $A_V$, $N$(\hi{}) and the metallicity suggest that the dust-to-metal ratios in GRB and QSO absorbers are roughly constant around the Galactic value \citep{Zafar13}.

An alternative approach to calculate the extragalactic dust-to-metal ratios is based on the dust-depletion pattern revealed by UV/optical absorption-line spectroscopy. Using such methods, the dust-to-metal ratios in QSO-DLAs and a few GRB-DLAs were found to be consistent with or slightly lower than the Galactic value \citep[][]{Vladilo98, Savaglio01, Savaglio03,Savaglio04,Vladilo04}. One of the advantages of using optical absorption lines rather than dust extinction to determine the dust content is that the absorption lines refer to a particular redshift and thus they are not integrated along the whole line of sight. This is particularly crucial for low-metallicity systems with an intrinsically low dust content and for high-$z$ targets, whose lines of sight likely cross intervening galaxies, each contributing to $A_V$.

In this paper, we develop a new dust-depletion-based method to derive the dust-to-metal ratios, cross-checked with the \citet[][S01 hereafter]{Savaglio01} method, and apply it to a sample of 20 GRB absorbers (observed at medium to high spectral resolution) and 72 QSO-DLAs (all with high-resolution spectroscopy). This approach solely relies on the observation of optical/UV low-ionization absorption lines associated with DLAs and therefore there is little or no ambiguity on the location or ionization state of the absorbing gas/dust: we purely focus on the bulk of the ISM in the host galaxies, which is mostly faraway from the GRB \citep[e.g., hundreds of parsecs from the burst][]{Vreeswijk07, Vreeswijk11} and therefore not affected by ionization due to the GRB or the progenitor. 

The paper is organized as follows. Our sample is defined in Sect. \ref{sample} and the methods of our analysis are explained in Sect. \ref{method}. The results are presented in Sect. \ref{results} and discussed in Sect. \ref{discussions}. Finally, we summarize our conclusions in Sect. \ref{conclusions}. Throughout the paper we adopt ions cm$^{-2}$ as the linear unit of column densities N. We refer to relative abundances of two chemical elements $X$ and $Y$ defined as $\left[X/Y\right] \equiv \log{\frac{N(X)}{N(Y)}} - \log{\frac{N(X)_\odot}{N(Y)_\odot}}$. For the reference solar abundances appearing in the second term of this formula, we use the values of \citet{Asplund09} and follow the recommendations of \citet{Lodders09} by adopting the photospheric estimates for the more volatile elements, the meteoritic estimates for the less refractory elements, or the average between them \citep[see details in][]{DeCia12}. The quoted errors and limits correspond to $1\,\sigma$ and $3\,\sigma$ significance levels, respectively, unless otherwise stated.

\section{The GRB-DLA sample}
\label{sample}

We select all GRB absorbers with a constrained estimate of the column density of \feii{}, and of at least one ion among \znii{} and \siii{}, from the largest sample to date of GRB absorber column densities collected by \citet{Schady11}. The resulting sample is composed of 20 GRB absorbers in the redshift range $1.2<z<4.0$ and spans a wide range of metallicities (from $Z\sim2\,Z_\odot$ down to $Z\sim0.005\,Z_\odot$), listed in Table \ref{tab dtm}. Most absorbers have an \hi{} estimate from \lya{}, all but one with $\log N$(\hi{}) $>20.3$ and therefore classified as DLAs \citep{Wolfe86}. One exception is GRB\,090323 with $\log N$(\hi{}) $=19.6$ for one of the two absorbing systems that composes its line profile. The column density estimates of the GRB sample are derived from absorption-line spectroscopy, using curve-of-growth analysis for the spectra with low or medium resolution \citep[see details in][and references therein]{Schady11} and Voigt profile fitting for a handful of high-resolution spectra \citep{Vreeswijk07,Ledoux09}.

\section{Revisiting the dust-to-metal ratio for GRB-DLAs: the \citet{Savaglio01} recipe and a new method.}
\label{method}

It is possible to estimate the dust-to-metal ratio from the depletion in dust of different elements, observed with optical/UV absorption-line spectroscopy, in particular from \feii{} and \znii{} lines. While \hi{} is fundamental to deriving the overall metallicity, it is not necessary to derive this ratio. \feii{} and \znii{} are two elements that have a similar nucleosynthetic history, but very different depletion properties. Fe is typically heavily depleted onto dust grains, while Zn is only mildly depleted \citep[e.g.,][]{Savage96,Ledoux02} and is therefore a metallicity indicator. Thus, we use [Fe/Zn] as a dust indicator, where the reliability of iron as a tracer of the bulk of the dust is explained in Appendix \ref{sect iron}. Si and S, also quite abundant in the ISM, can be useful dust-depletion indicators as well, but should be treated with caution, especially in the case of S \citep[e.g.,][]{Jenkins09}, as explained in Appendix \ref{sect sulphur}. Besides, the depletion of Si is typically not significant enough to make [S/Si] a solid dust indicator at low metallicity.

\subsection{The S01 method}

We derive the \dtm{} (referred to as $k/k_J$\footnote{$d/d_J$ in \citet{Savaglio03} and \citet{Savaglio04}} in S01) from the depletion pattern, i.e., from the observed abundance of different elements, as compared to the depletion pattern observed in different Galactic environments: halo (H), disk+halo (DH), warm disk (WD), and cool disk (CD), as listed in \citet{Savage96}. Following the recommendation of \citet{Jenkins09}, we update the mean depletion values listed in Table 6 of \citet{Savage96} slightly by adding $-0.03$, $-0.04$, and $-0.01$ to the values for Fe, Si, and S respectively, here and throughout the paper. The abundances with respect to H or Zn observed in GRB-DLAs are fitted simultaneously assuming that the Galactic depletion patterns, providing a best-fit \dtm{} and environment type. When both \hi{} and \znii{} are constrained, we treat the metallicity [Zn/H] as an input. Instead, when \hi{} is available but not Zn, we fit for the metallicity $Z/Z_\odot$ as well. In practice, we iterate among the different Galactic environments $J$ and sum over different metals $X$ to minimize the total difference between the observed abundances and the reference abundances
\begin{equation}
[X/\mbox{H}]_{\rm ref}=\log (1 + \mbox{\dtm{}}\cdot(10^{dXj} - 1) ) + \log (Z/Z_\odot) \mbox{,}
\label{eq S01}
\end{equation}
where $dXj$ is the depletion of a metal $X$ for a given environment $J$. More details can be found in S01 and \citet{Savaglio03}. We exclude upper and lower limits on the abundances from the fit. We include measurements for at least two metals $X$ among Zn, Si, and Fe. The results of the depletion pattern fit for each absorber are shown in the Appendix (Fig. \ref{fig: dtm_sav} and \ref{fig: dtm_sav_noH}, for the case with and without \hi{} constraints respectively). Sulphur is not included in the fit because it has shown controversial depletion patterns in different Galactic environments \citep{Jenkins09}, as described in Appendix \ref{sect sulphur}. The inclusion of silicon also must be taken with caution as the intrinsic abundance of Si with respect to Fe and Zn may not depend entirely on dust, but instead it could be boosted by $\alpha$-element enhancements in star-forming galaxies. This leads to a possible degeneracy between the best-fit \dtm{}, metallicity, and Galactic environments. We explore this possibility of an intrinsically lower Si by calculating the \dtm{}s for slightly lower Si column densities than observed ($\log N({\rm Si}) -0.2$ dex). The effect of such change is mostly accounted for by \textit{i)} setting an upper error on the \dtm{} of at least 0.2 for best-fit WD and DH environments, when Si is included in the fit and \textit{ii)} setting a lower error on the \dtm{} of at least 0.2 when Zn is not constrained.

\subsection{A new method for deriving the dust-to-metal ratio for DLAs}
\label{sec method}

    \begin{figure}
   \centering
   \includegraphics[width=90mm,angle=0]{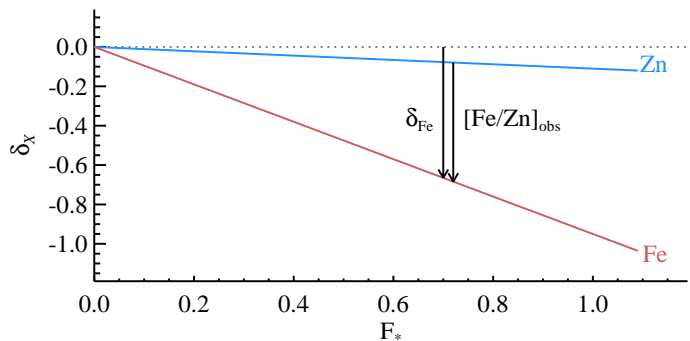}
   \caption{A scheme displaying two linear depletion sequences of Zn and Fe with the depletion strength factor $F_*$. Although $F_*$ is not known for a given line of sight, it can be traced by the observed [Zn/Fe] ratio.}
              \label{fig scheme}
    \end{figure}

Here we derive the absolute dust-to-metal ratio (\dtma{}) based on fundamental definitions and applying depletion corrections to the observed relative abundance of Fe with respect to Zn. The basic idea starts from the definition of the observed [Fe/Zn], where the observed gas-phase column density of an element $X$ is $N(X)_{\rm obs} = N(X)_{\rm tot}- N(X)_{\rm dust}$. If we assume that Zn is not depleted into dust and if Fe and Zn share the same nucleosynthesis (total of gas- and dust-phase [Fe/Zn]$_{\rm tot} = 0$), then a simplified version of the dust-to-metal ratio follows $\mbox{\dtma}_{\rm simplified}  =   1 - 10^{[\mbox{Fe}/\mbox{Zn}]_{\rm obs}} \mbox{.}$

To include the possible depletion of Zn\footnote{Some level of Zn depletion should be taken into account, given that GRBs are likely to reside in star-forming regions, some of them with a potentially significant dust content \citep[e.g.,][]{Perley13}.}, we assume that the depletion of an element in dust is proportional to a depletion strength factor $F_*$. This factor represents the overall strength of the dust depletion and it is unique to each line of sight, similar to what has been found for the Galaxy \citep[][e.g., their Fig. 7]{Jenkins09}. In the Galaxy, systems with little dust have $F_{*,\rm G}=0$ and almost no depletion of Zn, but the dustier systems $F_{*,\rm G}=1$ and even Zn is significantly depleted. We note that the $F_*$ scale for DLAs does not necessarily have to be the same as in the Galaxy because DLAs on average have lower metallicity and presumably lower dust contents. These linear depletion sequences reflect the tendency of each element to condensate in dust grains, i.e., they are related to the condensation temperature of each element \citep{Field74,Jenkins09}. Given that the slopes ($d$Fe, $d$Zn, $d$Si, et cetera) of these linear relations are specific to each element, the relative composition of  the dust grains changes depending on the amount of dust in the environment.

While the slopes and offsets of these linear depletion sequences have been well measured for the Galaxy \citep{Jenkins09}, in principle, we do not know how these relations change for lower-metallicity environments such as most DLAs. The observed offsets in the Galaxy are interpreted as pre-existing dust grains that cause some level of depletion even for the less dusty systems \citep{Jenkins09}. For the GRB and QSO sample of absorbers, we assume no offsets in the depletion sequences ($\delta_X=0$ at $F_*=0$), meaning that for DLAs we assume no pre-existing grains. This is expected given the relatively high redshift and the low metallicity of QSO- and GRB-DLAs, only a handful of which are above [Zn/H$]\sim-0.3$ in our DLA sample. Moreover, the evidence that most metal-poor DLAs (with assigned $F_*=0$) show [Zn/Fe] $\sim0$ \citep{Noterdaeme08,Molaro00} , i.e., no depletion, unambiguously reveals no offset in the DLA depletion sequences. \footnote{In the Galactic line-of-sight sample of \citet{Jenkins09}, the least dusty systems were not entirely dust-free, thereby leading to the observed Galactic offsets.}

In this way, the depletion of Zn and Fe can be described as 
\begin{subequations}
\begin{align}       
\delta_{\rm Zn} &= [\mbox{Zn/H}]_{\rm obs} - [\mbox{Zn/H}]_{\rm intrinsic} = d{\rm Zn} \cdot F_*    \label{eq linear Zn} \\               
 \delta_{\rm Fe} &= [\mbox{Fe/H}]_{\rm obs} - [\mbox{Fe/H}]_{\rm intrinsic} = d{\rm Fe} \cdot F_*\mbox{,}  \label{eq linear Fe}  
\end{align}       
\end{subequations}
where [Zn/H]$_{\rm intrinsic}$ and [Fe/H]$_{\rm intrinsic}$ are the total abundances in the gas and dust phases, and when $F_* = 1$, $\delta_{\rm Zn} = d{\rm Zn}$. Since the observed (depleted) abundances are associated with the gas-phase metals, the fraction of an element $X$ in gas $f_X$ is 
\begin{equation}        
 f_X = 10^{\delta_{X}}.
\end{equation}        
Figure \ref{fig scheme} visualizes these relations and the quantities we derive below. Note that we do not assume solar abundances or fixed grain compositions, rather we rely on the observed relative abundances [Fe/Zn]$_{\rm obs}$ as a function of the depletion strength. Since 
\begin{equation}
F_*= (\delta_{\rm Fe} -  \delta_{\rm Zn} )/(d{\rm Fe} - d{\rm Zn}) = \mbox{[Fe/Zn]}_{\rm obs}/(d{\rm Fe} - d{\rm Zn})\mbox{,}
\label{eq F*}
\end{equation}
the dust to metals ratio \dtma{} is then defined as
\begin{equation}
dtm = \frac{  N({\rm Fe})_{\rm dust} }{N({\rm Fe})_{\rm tot}} =1 - 10^{\delta_{Fe}} =  1 - 10^{  [{\rm Fe}/{\rm Zn}]_{\rm obs}\, d{\rm Fe}  /(d{\rm Fe} - d{\rm Zn} )  }  
\label{eq dtm}
\end{equation}
This definition of \dtma{} assumes only two linear dust depletion sequences for Fe and Zn respectively, and that Zn and Fe have the same nucleosynthetic history ($[{\rm Fe/Zn}]_{\rm tot}=0$), so any deviation of [Fe/Zn] from solar is due to dust depletion.

The slopes of the depletion sequences correspond to the expected level of depletion in the most dusty $(F_*=1)$ DLA systems. We assume $d{\rm Fe} = -0.95$, $d{\rm Si} = -0.26$  and $d{\rm Zn} = -0.11$, i.e., the average depletion along disk+halo lines of sight\footnote{Disk+halo lines of sight: HD 18100, HD 167756.} of Fe and Si \citep{Savage96} and Zn \citep{Roth95}. This translates to $\sim90$\% of Fe, $\sim 45$\% of Si, and $\sim 20$\% of Zn having been depleted into dust for the dustiest DLAs. The assumption of steeper slopes, such as the Galactic $d{\rm Fe} = -1.28$, $d{\rm Si} = -1.14$, and $d{\rm Zn} = -0.61$ \citep{Jenkins09}, would overestimate the dust-corrected metallicities (see Sect. \ref{discussions caveat} for a discussion on the dependence of our results on the assumed slopes). Thus, we limit our analysis to the slope values mentioned above and include an uncertainty of 0.1 on the slopes to allow for some freedom.

Finally, we normalize the \dtma{} values with respect to the Galactic value
\begin{equation}
  \mbox{\dtm{}} = dtm / dtm\mbox{(G),} 
  \end{equation}
where $dtm$(G) $=0.89$ (see the Appendix \ref{sect dtm_gal} for its derivation).

Another useful quantity that can be derived from the above formalism is the column density of iron in dust, $N({\rm Fe})_{\rm dust}$ (Col. 9 of Table \ref{tab dtm}), which provides an estimate of the dust content in the ISM, independently from $A_V$. Given that 
\begin{equation}
N({\rm Fe})_{\rm tot} =  N({\rm Zn})_{\rm tot} \cdot \left( \frac{N({\rm Fe})}{N({\rm Zn})}\right)_\odot 
\end{equation}
and
\begin{equation}
     N({\rm Zn})_{\rm tot} = \frac{N({\rm Zn})_{\rm obs}}{f_{\rm Zn}} =   \frac{N({\rm Zn})_{\rm obs}}{10^{{\rm[Zn/Fe]}_{\rm obs} \, d{\rm Zn} / ( d{\rm Zn} - d{\rm Fe} )}}\mbox{,} 
     \label{eq Zn tot}
\end{equation}     
the dust-phase column density of iron is
 \begin{align}
  N({\rm Fe})_{\rm dust} & =  N({\rm Fe})_{\rm tot} \cdot dtm  = \nonumber \\
   & = 10^{\rm[Fe/Zn]_{\rm obs}}  \cdot ( 10^{{\rm [Fe/Zn]}_{\rm obs} \, d{\rm Fe}/(d{\rm Zn} - d{\rm Fe}) } - 1 ) \nonumber \\   
  &\cdot N({\rm Zn})_{\rm obs} \cdot \left( \frac{N({\rm Fe})}{N({\rm Zn})}\right)_\odot \nonumber \\
  & = N({\rm Fe})_{\rm obs}  \cdot ( 10^{{\rm [Fe/Zn]}_{\rm obs} \, d{\rm Fe}/(d{\rm Zn} - d{\rm Fe}) } - 1 ) \mbox{.}
\label{eq fe dust}
\end{align}

   \begin{figure}
   \centering
   \includegraphics[width=90mm,angle=0]{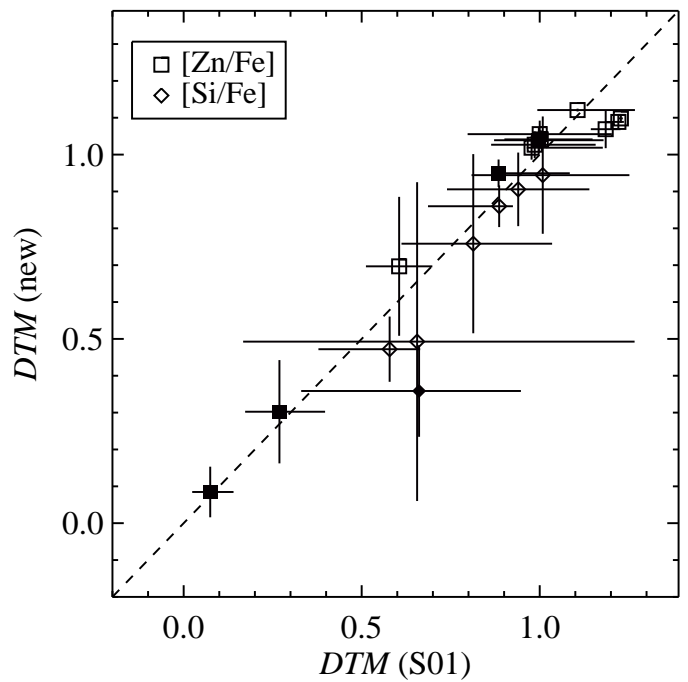}
   \caption{A comparison between the \dtm{} derived with the new and the S01 methods. The symbols indicate which metal was used as a reference to calculate the [$X$/Fe] ratio, where [Zn/Fe] is the most reliable ratio because it is less sensitive to intrinsic abundance changes. The filled symbols highlight the DLAs with high-resolution spectroscopy, which typically have more robust constraints. The vicinity to the equality (dashed) line shows the agreement between the two estimates.}
              \label{fig: dtm_comp}
    \end{figure}

\section{Results}
\label{results}

\subsection{Comparison among \dtm{} methods}

The \dtm{} derived with the new and the S01 methods for our GRB-DLA sample are listed in Table \ref{tab dtm} and plotted against each other in Fig. \ref{fig: dtm_comp}, for comparison. The two methods show a general good agreement and thus we consider them as equivalent (as also confirmed by a Spearman rank test\footnote{This indicates only how tight a monotonic correlation between two quantities is, but they do not necessarily need to be linear.}). The results of this test are listed in Table \ref{tab spearman}. There is a possible small deviation toward higher \dtm{}(S01) above 1, for three GRB absorbers, namely GBR\,010222, GRB\,050401, and GRB\,090323, absorbing system 2. We discuss whether such deviation could affect our results in Sect. \ref{discussions}. The overall reliability of both methods is strengthened by the agreement of the two different \dtm{} derivations because they are independent from each other.

In addition to the \dtm{} ratios, the S01 method also provides estimates of the metallicity in case the H column density is constrained but Zn is unconstrained, and the environment type \citep[halo, warm disk, cool disk, disk + halo, as characterized by][]{Savage96}. These are listed in the last two columns of Table \ref{tab dtm}. Note that when Si is constrained but Zn is unconstrained, the metallicity fitted from the depletion pattern is always higher than the observed [Si/H], as expected from the depletion of Si. Among the best-fit environments, a cool disk is somewhat disfavored for GRB absorbers, although a single best-fit environment could not be constrained for about half of our sample.

In the early stages of the current analysis we also derived the dust-to-metal ratio using the prescription of \citet{Vladilo98}, but we do not include those results because of the assumption of a constant grain composition (fixed fraction of each element in dust). Nevertheless, it is worth reporting that those dust-to-metal ratios are consistent with the results presented here.

\begin{landscape}
\begin{table}
\centering
\caption{dust-to-metal ratios normalized by the Galactic value (\dtm{}) derived with the method described in this paper and Savaglio's (2001, S01).}

\begin{tabular}{ l | r r r r r r>{\columncolor[gray]{ 0.9 }} r r r>{\columncolor[gray]{ 0.9 }} r  r r}
\hline \hline
\rule[-0.2cm]{0mm}{0.8cm}
GRB       &    $z$    & $A_V$ &  $\log$ & [$X$/H]$^a$& [$X$/Fe] &   $X$    &  \dtm{}            & [$X$/H]$_{\rm tot}^b$ & $\log$  $^c$ &  \dtm{}                & $\log Z/Z_\odot^d$   & $J^e$ \\ 
             &               &      (mag) &   $N(\mbox{\hi{}})$          &                 &                &              &  (new)  &   & \nfed{}             &  (S01) &                                &          \\
\hline
  & \\

   990123  &1.60 & $ <0.25             $ &                       &                           & $ 2.01\pm0.18   $&  Zn  &$ 1.12 \pm 0.00$& & $17.05 ^{+ 0.45}_{- 0.30}$&$ 1.11 ^{+ 0.16}_{- 0.11}$&                                &   all \\
   
   000926  &2.04 & $ 0.38 \pm 0.05$ & $21.30\pm0.20 $ & $-0.11\pm0.21 $ & $ 1.06\pm0.21   $&  Zn  &$ 1.06 \pm 0.04$&  $+0.03^{+0.34}_{-0.37}$ & $16.77 ^{+ 0.27}_{- 0.19}$&$ 1.00^{+0.20}_{-0.20}$&                                & D          \\
   
   010222  &1.48 & $ 0.24 ^{+ 0.08}_{- 0.09}$ &                       &                           & $ 1.30\pm0.17   $&  Zn  &$ 1.09 \pm 0.02$& & $16.78 ^{+ 0.33}_{- 0.23}$&$ 1.22 ^{+0.02}_{-0.02}$&                                &  H          \\
   
   020813  &1.25 &            &                       &                           & $ 0.90\pm0.07   $&  Zn  &$ 1.02 \pm 0.03$& & $16.45 ^{+ 0.25}_{- 0.18}$&$ 0.98 ^{+ 0.20}_{- 0.01}$&                                & D         \\
   
 030226$_{\rm s1}^f$  &1.99 & $ 0.05 \pm0.01$ & $20.50\pm0.30 $ & $-0.94\pm0.30 $ & $ 0.17\pm0.04   $&  Si  &$ 0.47 \pm 0.09$& $-0.87^{+0.34}_{-0.35}$ &  $14.72 ^{+ 0.19}_{- 0.15}$&$ 0.58 ^{+ 0.09}_{- 0.20}$&$-0.78 ^{+ 0.08}_{- 0.07}$ & DW       \\

 030226$_{\rm s2}^f$  &1.96 & $ 0.05 \pm0.01$ & $20.50\pm0.30 $ & $-1.04\pm0.30 $ & $ 0.45\pm0.03   $&  Si  &$ 0.86 \pm 0.06$& $-0.86^{+0.38}_{-0.39}$ & $14.99 ^{+ 0.24}_{- 0.15}$&$ 0.89 ^{+ 0.04}_{- 0.20}$&$-0.77^{+0.05}_{-0.05}$ & DW       \\

   030323  &3.37 &              & $21.90\pm0.07 $ & $-1.26\pm0.27 $ & $ 0.18\pm0.27   $&  Si  &$ 0.49 \pm0.43 $& $-1.19^{+0.25}_{-0.22}$ &$<16.46             $&$ 0.66 ^{+ 0.61}_{- 0.49}$&$-1.08 ^{+ 0.45}_{- 0.39}$ &  HDW    \\
   
   050401  &2.90 & $ 0.45 \pm0.03$ & $22.60\pm0.30 $ & $-0.93\pm0.42 $ & $ 1.14\pm0.36   $&  Zn  &$ 1.07 \pm0.05$& $-0.78^{+0.64}_{-0.73}$ & $17.27 ^{+ 0.53}_{- 0.45}$&$ 1.19^{+0.04}_{-0.04}$&                                &  H          \\
      
   050730  &3.97 & $ <0.17             $ & $22.10\pm0.10 $ & $-2.14\pm0.10 $ & $ 0.12\pm0.05   $&  Si  &$ 0.36 \pm0.12$&  $-2.09^{+0.12}_{-0.13}$ & $14.98^{+0.27}_{-0.27}$&$ 0.66 ^{+ 0.29}_{- 0.33}$&$-1.76 ^{+ 0.63}_{- 0.32}$ &  HWC    \\
   
     050820A  &2.62 & $ 0.27 \pm 0.04$ & $21.05\pm0.10 $ & $-0.72\pm0.10 $ & $ 0.98\pm0.12   $&  Zn  &$ 1.04 \pm0.03$&   $-0.59^{+0.21}_{-0.23}$ & $15.89 ^{+ 0.23}_{- 0.15}$&$ 1.00 ^{+ 0.18}_{- 0.13}$&                                &  HWC    \\

  050922C  &2.20 & $0.14 ^{+ 0.08}_{- 0.07}$ & $21.55\pm0.10$ & $-2.09\pm0.12$ & $ 0.35\pm0.27$&  Si &$0.76 \pm0.24$ &$-1.95^{+0.12}_{-0.16}$ & $14.89 ^{+ 0.31}_{- 0.58}$&$ 0.81 ^{+ 0.22}_{- 0.20}$&$-1.85 ^{+ 0.14}_{- 0.13}$ & DW       \\
    
   051111  &1.55 & $ 0.39 ^{+ 0.11}_{- 0.10}$ &                       &                           & $ 0.99\pm0.04   $&  Zn  &$ 1.04 \pm0.02$& & $16.41 ^{+ 0.25}_{- 0.17}$&$ 1.00 ^{+ 0.15}_{- 0.10}$&                                &  all \\
   
   060418  &1.49 & $ 0.13 ^{+ 0.01}_{- 0.02}$ &                       &                           & $ 0.71\pm0.03   $&  Zn  &$ 0.95 \pm0.04$& & $15.95 ^{+ 0.17}_{- 0.11}$&$ 0.88 ^{+ 0.20}_{- 0.02}$&                                & D          \\
   
   070802  &2.45 & $ 1.23 ^{+ 0.18}_{- 0.16}$ & $21.50\pm0.20 $ & $-0.16\pm0.24 $ & $ 0.37\pm0.18   $&  Zn  &$ 0.70 \pm0.19$&  $-0.11^{+0.34}_{-0.38}$ & $16.65 ^{+ 0.25}_{- 0.22}$&$ 0.61^{+0.09} _{-0.09}$&                                & C          \\   
   
   071031  &2.69 & $ <0.11             $ & $22.15\pm0.05 $ & $-1.76\pm0.05 $ & $ 0.03\pm0.03   $&  Zn  &$ 0.08 \pm0.07$& $-1.76^{+0.07}_{-0.07}$ & $14.74 ^{+ 0.09}_{- 1.26}$&$ 0.07 ^{+ 0.07}_{- 0.05}$&                                &  all \\
   
   080330  &1.51 & $ <0.19             $ &                       &                           & $ 0.93\pm0.11   $&  Zn  &$ 1.03 \pm0.04$& & $15.71 ^{+ 0.26}_{- 0.19}$&$ 0.99 ^{+ 0.17}_{- 0.12}$&                                &  all \\   
   
  080413A  &2.43 & $ <0.88             $ & $21.85\pm0.15 $ & $-1.63\pm0.16 $ & $ 0.12\pm0.07   $&  Zn  &$ 0.30 \pm0.14$& $-1.61^{+0.21}_{-0.22}$ &  $15.13 ^{+ 0.14}_{- 0.36}$&$ 0.27 ^{+ 0.13}_{- 0.10}$&                                &  HWC    \\ 
 
 090323$_{\rm s1}^f$  &3.58 & $ 0.10 \pm 0.04$ & $19.62\pm0.33 $ & $ +0.33\pm0.35 $ & $ 0.51\pm0.14   $&  Si  &$ 0.91 \pm0.10$& $+0.53^{+0.46}_{-0.50}$ & $15.52 ^{+ 0.35}_{- 0.26}$&$ 0.94 ^{+0.20}_{-0.20}$&$ 0.60^{+0.16}_{-0.16}$ & D          \\ 
 
 090323$_{\rm s2}^f$  &3.57 & $ 0.10 \pm 0.04$ & $20.72\pm0.09 $ & $ +0.22\pm0.10 $ & $ 1.41\pm0.06   $&  Zn  &$ 1.10\pm0.01$& $+0.40^{+0.28}_{-0.30}$ & $16.58 ^{+ 0.32}_{- 0.22}$&$ 1.23^{+0.01} _{-0.01}$&                                &  H          \\
 
  090926A  &2.11 & $ <0.15             $ & $21.73\pm0.07 $ & $-2.27\pm0.27 $ & $ 0.57\pm0.30   $&  Si  &$ 0.94 \pm0.16$&  $-2.05^{+0.25}_{-0.34}$ & $15.07 ^{+ 0.50}_{- 0.40}$&$ 1.01 ^{+ 0.24}_{- 0.20}$&$-1.94 ^{+ 0.38}_{- 0.35}$ &  HDW    \\
  & \\
  \hline \hline
 \end{tabular}
 \tablefoot{Uncertainties refer to a $1~\sigma$ confidence level, while $3~\sigma$ limits are reported for unconstrained measurements. $^a$ Observed abundance of the element $X$ relative to solar. $^b$ Abundances corrected for dust depletion. When available, [Zn/H]$_{\rm tot}$ is considered to be the metallicity of the gas. $^c$ Column density of iron in dust-phase (see text). $^d$ Metallicity as fitted with the S01 method (Eq. \ref{eq S01}), when H is unconstrained and Zn constrained. $^e$ Galactic environments fitted with the S01 method: halo (H), disk+halo (D), warm disk (W) and cool disk (C), as characterized by \citet{Savage96}. When more than one environment is possible, \dtm{}(S01) and $\log Z/Z_\odot$ are averaged among the allowed values. $^f$ Different absorbing systems toward the same GRB are reported separately.}
\tablebib{All column densities, $z$ and $A_V$ are taken from the compilation by \citet{Schady11}.}
\label{tab dtm}
\end{table}

\end{landscape}

\subsection{Comparison with QSO-DLAs}

To compare the GRB- and QSO-DLA populations, we calculate the \dtm{} for a sample of 72 QSO-DLAs from \citet{Noterdaeme08}, all of which have been observed with the high-resolution Very Large Telescope (VLT) Ultraviolet and Visual Echelle Spectrograph \citep[UVES,][]{Dekker00}. We choose to compute the \dtm{} of QSO-DLAs with our new method because it is easily derived from the [$X$/Fe] reported in \citet{Noterdaeme08}. The overall sample of GRB- and QSO-DLAs spans a wide range of metallicities, from $Z\sim2\,Z_\odot$ down to $Z\sim0.002\,Z_\odot$. The average observed [Zn/Fe] in the QSO and GRB samples are 0.4 and 0.9, respectively, with a standard deviation of 0.2 and 0.5. This suggests that GRB environments are more dusty overall than QSO-DLAs. Indeed, only less than $\sim10$\% of QSO-DLAs show significant reddening due to dust \citep{Khare12}.

Figure \ref{fig: dtm_distr} shows the distribution of \dtm{}s, derived with the new method, in GRB-DLAs (black) and QSO-DLAs (grey) as a function of metallicity (left panel) and the Zn column density (right panel) corrected for dust depletion using Eq. \ref{eq Zn tot}. The depletion corrections are typically very small, up to $\sim$0.3 dex. In Fig. \ref{fig: dtm_distr with Si}, we also include those targets that have measurements of the Si column density but lack measurements of the Zn column density. The \dtm{} that has been estimated from Zn should be considered more reliable because the [Zn/Fe] is less affected by intrinsic changes of relative abundances than [Si/Fe].

\subsection{Comparison with the Magellanic Clouds}

\begin{figure*}
   \centering
   \includegraphics[width=170mm,angle=0]{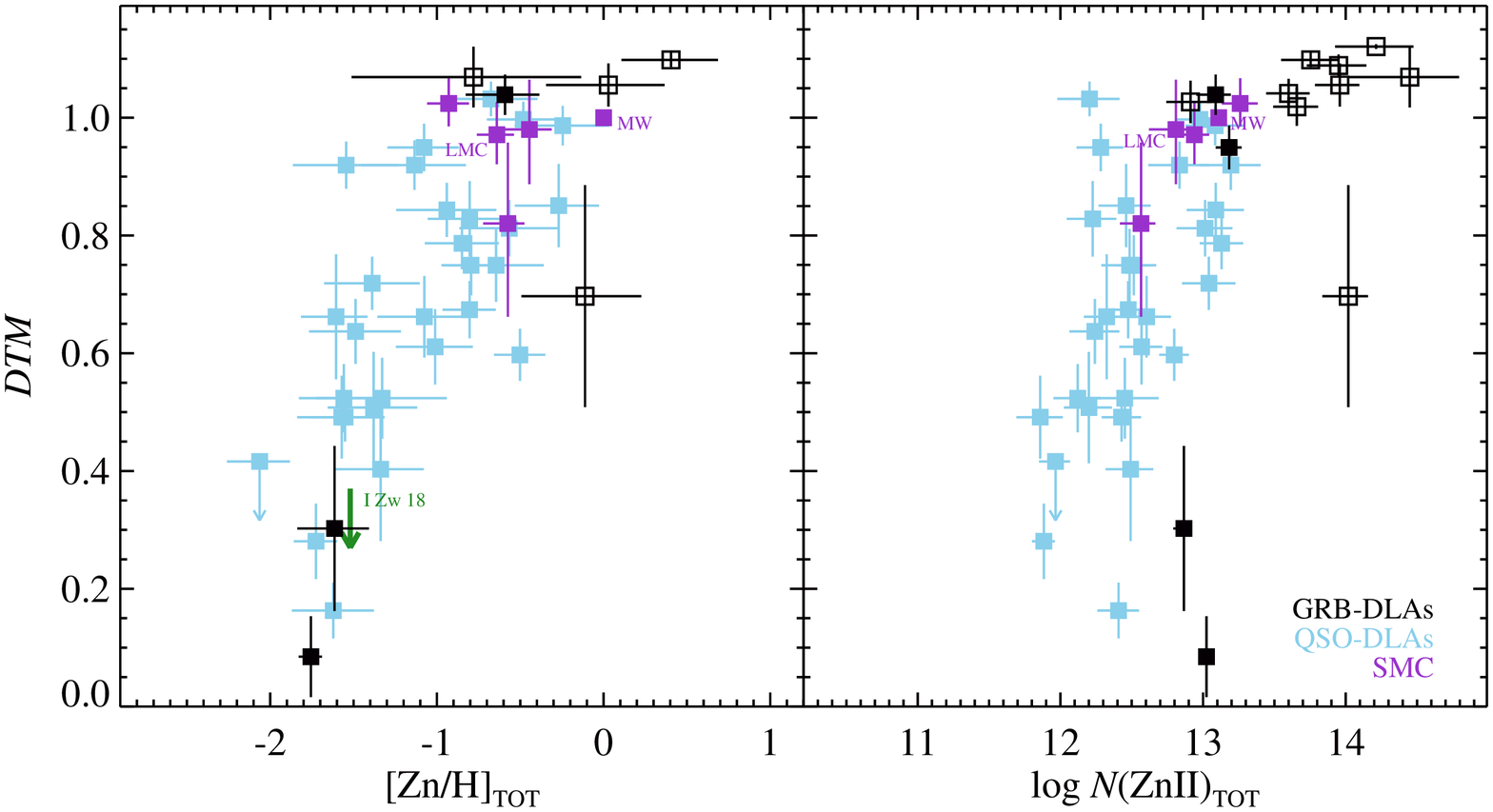}
   \caption{\textit{Left:} The distribution of the \dtm{} with dust-corrected [Zn/H]$_{\rm TOT}$ (the intrinsic metallicity), for the GRB-DLAs (black) in our sample and the UVES QSO-DLAs (gray) sample of \citet{Noterdaeme08}. The MW, SMC (three lines of sight) and LMC \dtm{} values are displayed for comparison, where the latter two are two representative values derived from the observed [Zn/Fe] along two lines of sight (see text for details). Filled symbols highlight the measurements drawn from high-resolution spectroscopy. The thick green arrow shows the literature value \citep[derived from the $M_{\rm dust}/M_{\rm H}$ measured by][]{HerreraCamus12} for the nearby blue compact dwarf galaxy I Zw 18. \textit{Right:} The distribution of the \dtm{} with the total column density of \znii{}, corrected for dust depletion.}
              \label{fig: dtm_distr}
   \includegraphics[width=170mm,angle=0]{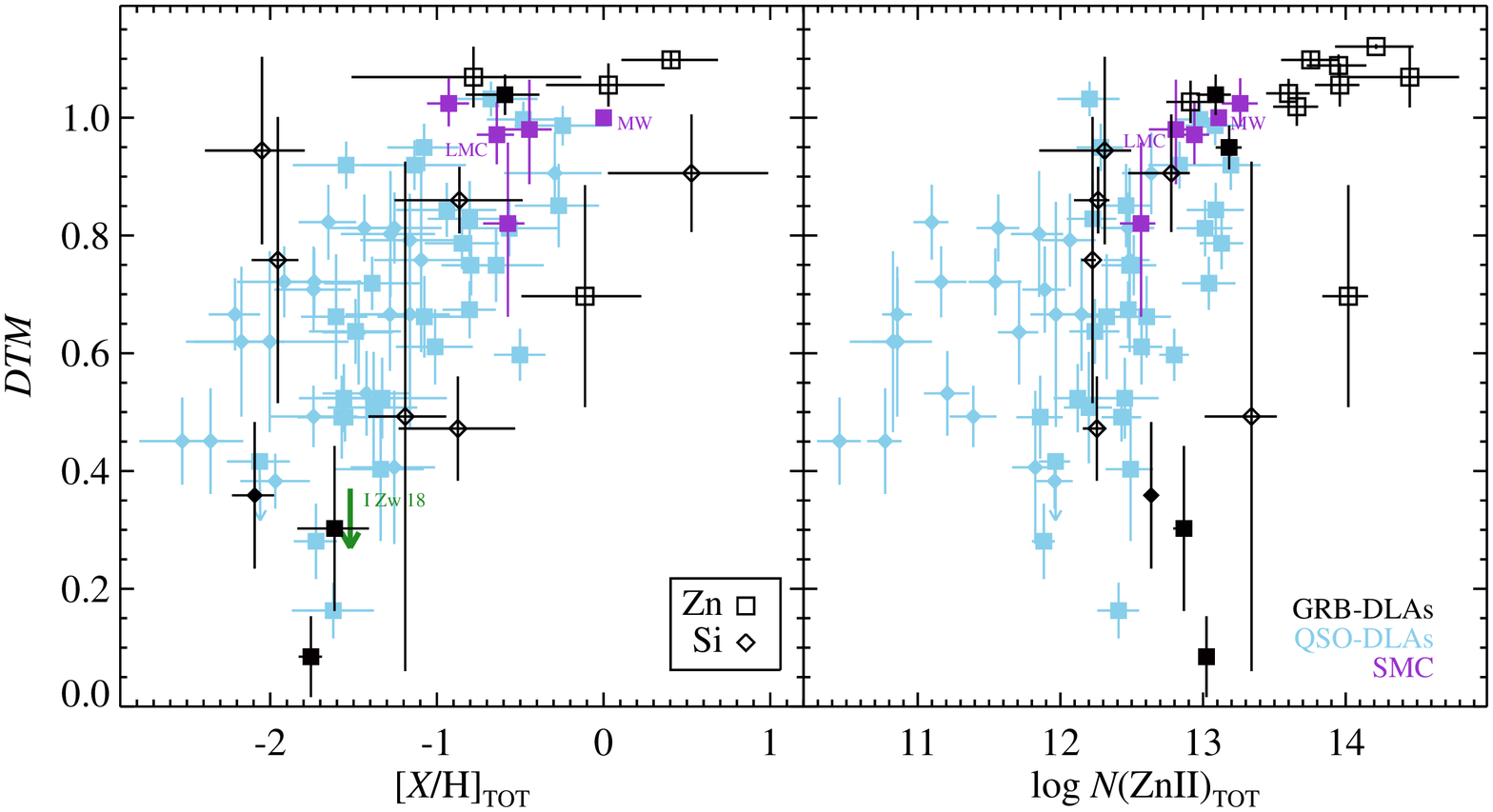}
   \caption{Same as Fig. \ref{fig: dtm_distr}, but including the \dtm{} values that have been estimated from Si and thus are less reliable. When Zn is not directly constrained, we display the equivalent \znii{} column density derived from the \siii{} column density, assuming solar relative abundances. The symbols indicate which metals are used in [$X$/Fe] to derive the \dtm{}.}
              \label{fig: dtm_distr with Si}
    \end{figure*}

We calculate the \dtm{} values for two lines of sight through the SMC and the LMC, using the Zn and Fe column densities derived from the absorption lines observed in the literature. For the SMC, we consider the absorption in spectra of the background star Sk 108 \citep{Welty97}\footnote{We refer to the mean observed [Zn/Fe] $=0.50\pm12$ and [Zn/H] $=-0.64^{+0.13}_{-0.17}$ associated with the SMC, consistent with [Zn/Fe] $=0.66\pm20$ measured by \citet{Sofia06} for the same line of sight.}, Sk 155, and Sk 78 \citep{Welty01}. For the LMC, we use the absorption toward SN 1987A \citep{Welty99}\footnote{We refer to the mean [Zn/Fe] $=0.78^{+0.12}_{-0.16}$ and [Zn/H]$=-0.55\pm0.08$, averaged among the velocity components associated with the LMC.}. While the choice of a few specific line of sight per system may not represent the overall properties of the whole cloud, we note that typically only one line of sight can be observed toward both GRB and QSO as well. The derived \dtm{} are $\sim$82--100\% and $\sim$98\% of the Galactic value for the SMC and the LMC, respectively, as displayed in Figs. \ref{fig: dtm_distr} and \ref{fig: dtm_distr with Si}. Given the similar [Zn/H] and $\log N$(Zn) of the SMC and LMC absorbers, it is interesting to note that the LMC has a higher [Zn/Fe], \dtm{}, and ultimately a larger amount of dust with respect to SMC, although these differences are not $3~\sigma$ significant.

\section{Discussion}
\label{discussions}

\subsection{Strengths, caveats, and implications of our \dtm{} method}
\label{discussions caveat}
The estimates of the dust-to-metal ratio presented in this paper rely purely on the metal absorption lines observed in the optical/UV afterglow spectra. These are selected at the GRB redshift, typically arising in the ISM of the host galaxy, hundreds of parsecs from the burst \citep[e.g.,][]{Vreeswijk07,Vreeswijk11}.

The main assumption of our \dtm{} method is the existence of linear depletion sequences in DLAs similar to what has been observed in the Galaxy \citep{Jenkins09}. These sequences depend on the condensation temperature of each chemical element and naturally explain different dust-grain compositions in different environments. While the slopes of the Galactic depletion sequences are well studied \citep{Jenkins09}, the corresponding slopes in DLAs are unknown. Here we discuss how much our results depend on the assumption of these slopes.

At the low metallicity end, i.e., in the least dusty DLAs, the low depletions are not very sensitive to changes in slopes (see Fig. \ref{fig scheme}), while the zero offsets of the linear sequences are fully justified by the observation of [Zn/Fe] $\sim0$ in low-metallicity systems (see also Sect. \ref{sec method}). For the most dusty DLA systems, the dependence on the slopes of the depletion sequences are more significant. In comparison with the Galactic values however, we find that the possible range of slopes is limited. The slopes of the depletion sequences for DLAs must be shallower than in the Galaxy to provide reasonable dust-corrected metallicities in the DLAs. Indeed, using the Galactic values of the depletion-sequences slopes generally gives a small increase of \dtm{} (up to $\sim0.15$), but implies excessive metallicities. In the most extreme cases, the Galactic slopes lead to dust-corrected metallicities of up to $Z\sim7\,Z_\odot$ and $Z\sim120\,Z_\odot$ for GRB\,000926, and GRB\,090926A, which are individual absorbing systems with [Zn/Fe]- and [Si/Fe]- based \dtm{} respectively. These values are too large to be realistic. These numbers are even more extreme for dustier absorbers such as GRB\,090323, where the Galactic slopes lead to dust-corrected $Z\sim30\,Z_\odot$ and $Z\sim10^4\,Z_\odot$ for the two separate absorbing systems. These results are not justifiable, regardless of the degeneracy of the \hi{} content between the two absorbing systems. On the other hand, the slopes that we assumed for DLAs provide dust-corrected metallicities $Z\sim1.1\,Z_\odot$, $Z\sim0.01\,Z_\odot$, $Z\sim3.4\,Z_\odot$, and $Z\sim2.5\,Z_\odot$ for the systems mentioned above, respectively. Thus, the depletion levels of Si, Fe, and Zn are likely more moderate in DLAs than in the Galaxy. The \dtm{} values given in Table \ref{tab dtm} are probably reliable within the errors, despite the uncertainties in the slopes of the depletion sequences in DLAs. 
 
The depletion of Fe and Zn of the most dusty DLAs (with depletion factor $F_*\sim1$ and above) are similar to what has been observed in the least dusty lines of sight in the Galaxy \citep[with Galactic depletion factor $F_{*,\rm G}\sim0$,][]{Jenkins09}, such as disk+halo lines of sight studied by \citet{Savage96}. One intriguing possibility is that the depletion factor could evolve continuously from the least depleted dust-free DLAs to the most dusty Galactic lines of sight, where the depletion sequences would start at $F_*=0$ with no offset and shallow slopes and gradually evolve into the observed Galactic depletion sequences.

\subsection{The \dtm{} distribution}

\textit{The \dtm{} distribution of the GRB-DLAs in our sample increases with metallicity and spans $\sim$10\%--110\% of the Galactic value.  In addition, the observed trend points to low dust fractions for low-metallicity systems, which is inconsistent with a constant Galactic \dtm{} at any metallicity, as recently suggested \citep[e.g.,][]{Zafar13}. Given these results, it is clear that \avnh{} should not be considered a reliable estimate of the dust-to-metal ratios in GRB-DLAs.} 

The comparison between GRB- and QSO-DLAs in Fig. \ref{fig: dtm_distr} reveals a continuous distribution of the \dtm{} between the two populations, indicating an overall similarity of their ISMs. The mean \dtm{} for GRBs and QSOs are 0.8 and 0.6 respectively, with a standard deviation of 0.3 and 0.2. Given that GRB lines of sight preferentially cross through inner (and more dusty) regions of their star-forming galaxies, while QSO lines of sight typically intercept the haloes of intervening galaxies \citep[e.g.,][]{Prochaska07, Fynbo09}, GRBs probe the \dtm{} distribution up to one order of magnitude larger column densities than QSO lines of sight. In the Galaxy, the average volume density of the gas is tightly correlated with the depletion factor \citep[$F_{\rm *,G}=0.772+0461\cdot\log\langle n({\rm H})\rangle$][]{Jenkins09}. For DLAs, we derive the average volume density from the observed the depletion factor $F_*$ (Eq. \ref{eq F*}), using the relation observed in the Galaxy. While the actual values of $\langle n({\rm H})\rangle$ that we find for DLAs may be unreliable because the relation might evolve with metallicity and be different for DLAs, the comparison between the densities in QSO- and GRB-DLAs are meaningful. While QSO-DLAs have $0.001<\langle n({\rm H}) \rangle<10$ cm$^{-3}$, several GRB-DLAs exceed $10$ cm$^{-3}$, confirming that GRBs are typically located in denser regions than QSO absorbers.

The \dtm{} increases with metallicity (Fig. \ref{fig: dtm_distr} and \ref{fig: dtm_distr with Si}, left panels). Despite the large scatter, the increasing trend is clear up to [$X$/H$] \sim-1$. Above this metallicity the distribution may be consistent with a roughly constant dust-to-metal ratio for DLAs, as found by \citet{Zafar13} based on $A_V$. At lower metallicities we find low \dtm{} values in both QSO- and GRB-DLAs, which are not constant and not consistent with the Local Group, pointing to a non universal dust-to-metal ratio. 

The discrepancy between our results and the $A_V$-derived \dtm{} \citep[][although only one single datapoint is constrained at very low metallicities]{Zafar13} might arise from an overestimate of the intrinsic dust content when estimated from the line-of-sight $A_V$ of the lower-metallicity and less dusty systems. For these low-dust systems the additional extinction due to foreground absorbers along the whole line of sight contribute significantly to the $A_V$. Indeed, the observed foreground $E(B-V)$ toward QSOs typically ranges between $\sim$0.001 and $\sim$0.1 \citep{Menard08}, i.e., $0.003 \lesssim A_V \lesssim 0.3$, so this additional source of dust extinction can be significant when compared to the intrinsically low $A_V$ of very low-metallicity systems. On the other hand, our dust-depletion analysis does not suffer from the contribution of intervening systems and therefore provides reliable estimates of the dust content of low-metallicity systems. 

In Figs. \ref{fig: dtm_distr} and \ref{fig: dtm_distr with Si}, we include a literature estimate of the \dtm{} of the blue compact dwarf galaxy I Zw 18, one of the lowest-metallicity systems known in the local universe \citep[{$\left[{\rm O/H}\right] =-1.52$}, $M_{\rm dust}/M_{\rm H}<8.1\times 10^{-5}$;][]{HerreraCamus12}, where we consider the dust-to-metal mass ratio $M_{\rm dust}/M_{\rm met}=10^{\log (M_{\rm dust}/M_{\rm H}) - {\rm [O/H]}}$ and normalize it by the Galactic value $(M_{\rm dust}/M_{\rm H})_{\rm Gal}=0.0073$ \citep{Draine07b}. The \dtm{} $<0.37$ of I Zw 18 is consistent with the trend that we find for GRB- and QSO-DLAs, supporting the idea that low metallicity systems have lower dust fractions than typical spiral galaxies.

The \dtm{} distribution increases with the metal column density as well (represented by the Zn or the equivalent Zn column density, right panel of Fig. \ref{fig: dtm_distr}) and flattens out above $\log N(\mbox{Zn})\sim13$. One issue that needs to be clarified is whether this flattening is physical or is artificially introduced by our analysis. As noted in Sect. \ref{results}, our method might slightly underestimate three \dtm{} values (see Fig. \ref{fig: dtm_comp}), all of which have $\log N(\mbox{Zn})>13$. However, given the magnitude of the effect, these three points are unlikely to prove crucial. The observed flattening of the \dtm{} at higher metal column densities might suggest that the availability of metals in the ISM can drive dust formation/growth up to some stable level where metals are largely accessible, as is the case for the Galaxy. Dust destruction due to GRBs might also play a role in limiting the amount of dust \citep[e.g.,][]{Waxman00,Draine02}.

The significance of the correlations described above is confirmed by a Spearman rank test, as listed in Table \ref{tab spearman}, indicating that \dtm{} is most tightly correlated with [Zn/H]. There is a much weaker correlation between the \dtm{} distribution and the \hi{} column density, as shown in Fig. \ref{fig: dtm_HI} in the Appendix. This confirms that the dust traces the metals rather than the total gas content.

While the \dtm{} values in QSOs are uniformly distributed, we cannot exclude the presence of two populations of GRB hosts divided by their \dtm{} properties, i.e., a standard one following the trends that have been observed for QSO-DLAs and a second group of GRB-DLAs with a lower \dtm{}. This is most notable in the \dtm{} distribution against the metal column density, although the gap is less evident when also including \dtm{} estimates derived from Si. It is difficult to assess whether this separation is a real effect, given the small number statistics. If true, these low-\dtm{} GRB-DLAs (mostly GRB\,071031 and GRB\,080413A) may correspond to systems having a smaller amount of dust for their metal content, possibly due to sputtering or sublimation of dust grains in actively star-forming regions \citep{Tielens94,Waxman00}. Further observations are clearly needed to investigate this issue.

We do not observe any trend of the \dtm{} with the mass, star-formation rate, or specific star-formation rate of the host galaxies, although the sample for which all these estimates are available is limited to a handful of objects \citep{Vreeswijk04,Chen09,Savaglio09,Kruehler11,Savaglio12}.

\begin{table}
\centering
\caption{Spearman's rank correlation coefficients $\rho$ and significance.} 
\begin{tabular}{@{}l@{\hspace{3mm}} l | l@{\hspace{3mm}} r@{\hspace{3mm}} r@{\hspace{3mm}} r@{\hspace{3mm}} r@{}}
\hline \hline
\rule[-0.2cm]{0mm}{0.8cm}

$x^a$ & $y^a$    & Fig. & DLA & $\rho^b$   & $P^c$ & $\sigma$ \\  
\hline
\dtm{}(S01) & \dtm{} & \ref{fig: dtm_comp} & GRB &   0.94 & $<0.001$  & 6.15 \\ 
 
 \hline            
 \multirow{2}{*}{[Zn/H]$_{\rm TOT}$} & \multirow{2}{*}{\dtm{}}  & \multirow{2}{*}{\ref{fig: dtm_distr}}     & QSO &  0.660 &  $<0.001$ & 3.8\\
                                                                                                     &         &           & GRB &  0.667  &        0.050 & 1.96\\ 

   \hline
\multirow{2}{*}{$\log N(\mbox{\znii{}})_{\rm TOT}$} & \multirow{2}{*}{\dtm{}}  & \multirow{2}{*}{\ref{fig: dtm_distr}}   & QSO &  0.464 &  0.013 & 2.48  \\
                                                                                                                              &          &         & GRB &  0.632  &   0.020 & 2.32\\

\hline
\multirow{2}{*}{[$X$/H]$_{\rm TOT}^d$} & \multirow{2}{*}{\dtm{}}  & \multirow{2}{*}{\ref{fig: dtm_distr with Si}}     & QSO &  0.564 &  $<0.001$ & 4.9 \\
                                                                                                     &         &           & GRB &  0.604  &        0.022 & 2.29\\

  \hline        
\multirow{2}{*}{$\log N(\mbox{\znii{}})_{\rm TOT, Eq}^e$} & \multirow{2}{*}{\dtm{}}  & \multirow{2}{*}{\ref{fig: dtm_distr with Si}}   & QSO &  0.265 &  0.062 & 1.86\\
                                            &         &         & GRB &  0.682  &   $<0.001$ & 3.31\\

\hline
\multirow{2}{*}{$z$} &  \multirow{2}{*}{\dtm{}}  & \multirow{2}{*}{\ref{fig: dtm_z}}   &  QSO &  $-0.052$ &  0.671 & 0.43\\
                                                                                                &      &        & GRB &  $-$0.338  &   0.144  & 1.46\\

\hline                                                                                                

 \nfed{}$_{\rm, Zn, Si}$  & \multirow{5}{*}{$A_V$} & \multirow{5}{*}{\ref{fig: Av_NFedust}}  & \multirow{5}{*}{GRB} &  0.702  &  0.011 & 2.54 \\

 \nfed{}$_{\rm, Zn}$  & & & &  0.357  &  0.385  & 0.88\\
 \nfed{}$_{\rm, Si}$  & & & &  0.305  &  0.392  & 0.86\\
 \nfed{}$_{\rm, Zn}^f$& & & &  0.357  &  0.431  & 0.79\\
 \nfed{}$_{\rm, Si}^f$ & & & &  0.798  &  0.010  & 2.58\\

\hline 

\multirow{2}{*}{$N$(\hi{})} & \multirow{2}{*}{\dtm{}}  & \multirow{2}{*}{\ref{fig: dtm_HI}}   &   QSO &  $-0.123$ &  0.318 & 1.00\\
                                                                                                             &      &        & GRB &  $-$0.513  &  0.021 & 2.31\\

  \hline \hline
 \end{tabular}
\tablefoot{$^a$ $x$- and $y$-axis sample populations for the correlation test. $^b$ $-1 \leq \rho \leq +1$, where $\rho=+1(-1)$ for perfect monotonic increasing (decreasing) correlations. $^c$ Null-hypothesis probability, where the number of $\sigma$ indicates the level of significance of the correlation. $^d$ $X$ is either Zn or Si. $^e$ Zn or equivalent Zn column density derived from Si, see Fig. \ref{fig: dtm_distr with Si}. $^f$ Excluding the outlier GRB\,070802}
\label{tab spearman}
\end{table}

\subsection{On the origin of dust in DLAs}

\begin{figure}
   \centering
   \includegraphics[width=90mm,angle=0]{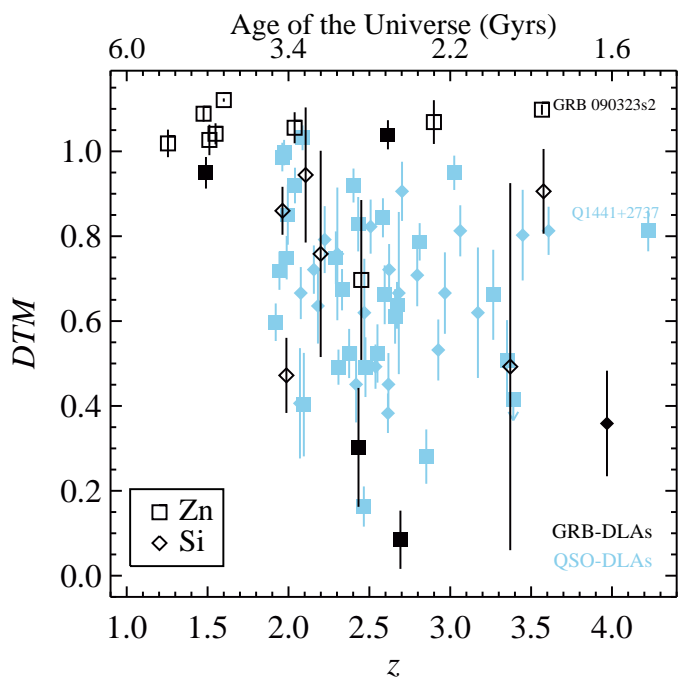}
   \caption{The \dtm{} evolution with redshift. The symbols are the same as in Fig. \ref{fig: dtm_distr with Si}.}
              \label{fig: dtm_z}
    \end{figure}

We further explore the evolution of the \dtm{} with redshift, as displayed in Fig. \ref{fig: dtm_z}. Combined with the metallicity information, this provides some further clues on the dust production in DLAs.

\subsubsection{At low redshift} 
All the estimated \dtm{} values for GRB-DLAs below $z\sim1.7$ (6 out of 20) are large and similar to the Galactic value. While the metallicity of these systems has not been measured from absorption lines because Ly-$\alpha$ is not redshifted into the observable window, the \znii{} column densities of these absorbers are all above $\log N$(\znii{}) $\gtrsim13$ \citep[with the exception of GRB\,080330 with $\log N$(\znii{}) $\sim12.8$][]{D'Elia09b}. Low-$z$ GRB absorbers follow the \dtm{} distribution with the metal column density discussed above and it is reasonable to expect that they follow the \dtm{} distribution with metallicity as well. If so, this would indicate that at redshift below $z\lesssim1.7$ most GRBs occur in galaxies with metallicities $Z\gtrsim0.1Z_\odot$, possibly because the mean metallicity of star-forming galaxies has increased with cosmic evolution. However, this result should be viewed with caution because the number of observations at low redshift is limited.
 
\subsubsection{At high redshift} 
While the \dtm{} distribution shows a large scatter above $z\gtrsim1.7$, we confirm the evidence for some high-\dtm{} galaxies, such as GRB\,090323 at $z=3.6$ \citep{Savaglio12} or QSO-DLA Q 1441+2737 at $z=4.2$ \citep{Ledoux06}, to lie at high redshift. The presence of dusty systems at these redshifts suggests that a rapid dust-production mechanism may be required for the dust to have formed well within a Gyr. One such fast mechanism is the condensation of dust grains in cooled SN ejecta \citep[e.g.,][]{Matsuura11}, where the contribution of SNe on dust production is dominant at early times \citep[for time scales below $\sim200$ Myrs,][]{Gall11}. AGB stars start to recycle their metals into the ISM producing silicates grains after $\sim40$ Myrs \citep{DiCriscienzo13} and carbon-rich dust after a few hundreds Myrs of main-sequence evolution \citep{Galliano08}. However, at low metallicities the contribution of AGB stars to dust production is unlikely to be significant in systems with young stellar populations (see below). The observation of rapidly-produced dust (within 30--170 Myrs) in $z\gtrsim6$ QSOs suggests that SNe dominate the dust production at these redshifts, with a negligible contribution of AGB stars \citep{Gall11b}. While SNe are the dominant source of dust at early times, AGB stars are thought to start contributing at $z <$ 8--10 \citep{Valiante11}. However, extinction curves of SN-produced dust (showing a kink at 3000 \AA{}) have not been conclusively constrained in $z\sim6$ quasars \citep{Hjorth13} yet. 

Grain growth in the ISM is typically considered a slow process for producing dust, but its time scale decreases with increasing metallicity \citep[the growth time scale $\tau$ is of the order of $\sim$60 Myrs for solar-metallicity systems,][]{Hirashita11}. Thus, dust production via grain growth cannot be excluded for those systems at high redshift with relatively high metallicity, such as GRB\,090323,  but also down to a few percent solar metallicity. Moreover, ISM shattering can increase the number of small grain and thereby allow grain growth to account for the amounts of dust observed in high-redshift quasars \citep{Kuo12}. Besides, grain growth is required to explain the dust content observed in high-redshift quasars \citep{Mattsson11,Valiante11}. While the current observation of the \dtm{} at $z\sim4$ cannot put stringent limits on the origin of this dust, future observations of \dtm{} at redshifts above 5 will be crucial to assess the extent of SN contribution at high redshift \citep{Morgan03}. The method described in this paper can in principle allow the \dtm{} to be measured up to very high redshift. 

\subsubsection{At low metallicities} 
The low \dtm{} of $Z\lesssim0.1Z_\odot$ DLAs implies that these systems have a lower fraction of dust with respect to the higher metallicity systems. This could be due to a lower efficiency of dust production and/or a higher efficiency of dust destruction. The efficiency of dust-production from AGB stars is also lower in low-metallicity systems, becoming a negligible source of dust below $Z\sim0.005Z_\odot$ \citep{DiCriscienzo13}. However, in low-metallicity environments the most massive AGB stars that are short-lived, but not massive enough to explode as SNe ($5M_\odot\lesssim M \lesssim 8M_\odot$) evolve to become N-rich and O-rich M-stars \citep{Karakas07,Gall11c} that are particularly inefficient at forming dust \citep{Ventura12}. Thus, AGB stars seem to bring a negligible contribution to dust production, when compared to SNe, in low-metallicity systems dominated by young stellar populations. Importantly, SNe dust production is not very sensitive to metallicity changes \citep{Galliano08}, and thus represents an unlikely source of the \dtm{} trend with metallicity. On the other hand, below $Z\sim0.1 Z_\odot$ grain growth starts to be less efficient and its time scale becomes larger \citep[$\tau$ of hundreds of Myrs to Gyr, depending on the metallicity and the grain-size distribution;][]{Hirashita11}.

One possible cause of a higher efficiency in dust destruction could be a harder and more intense radiation field produced by enhanced massive-star formation in low-metallicity environments, such as for GRB host galaxies \citep[e.g.][]{Savaglio09}. However, \citet{Gall11} showed that simulated dust masses are higher for initial mass functions with more high-mass stars, meaning that SNe are overall more efficient at producing than destroying dust. An alternative cause of a higher level of dust destruction could be a grain-size distribution skewed toward smaller grains in low-metallicity systems \citep[][]{Sandstrom12}, since such dust grains are more easily destroyed \citep[e.g.,][]{Waxman00}. However, this is an unlikely scenario for QSO-DLAs, where sources of dust destruction (shocks or radiation due to SNe or GRBs) are limited. Moreover, an increasing \dtm{} with increasing metallicity is expected only if there is a net production of dust via grain growth in the ISM over dust destruction \citep{Mattsson12}\footnote{\dtm{} gradients that are coupled with metallicity gradients are thought to be a signature of the higher efficiency of grain growth over dust destruction \citep{Mattsson12}.}. \textit{Thus, the lower fraction of dust in the low-metallicity DLAs could be due to a lower efficiency of dust-production via grain growth, with negligible SN and AGB stars contribution.}

\subsection{Dust extinction from the host galaxy ISM}

Given that the measurement of the dust extinction $A_V$ is integrated along the whole line of sight, it is not trivial to assess whether the intrinsic $A_V$ of the host galaxy is mostly associated with the bulk of the ISM. In principle, there could be a contribution from intervening systems \citep[although they show typically weaker absorption than GRB-DLAs themselves;][]{Vergani09} or from a dense and dusty star-forming region associated with the GRB. On the other hand, we use the metal absorption lines to infer the dust column in the host galaxy ISM, specifically the column density of Fe in dust \nfed{} (Eq. \ref{eq fe dust}). It is insightful to compare $A_V$ and \nfed{} because they are two entirely independent estimates of the dust content: $A_V$ is derived from the continuum shape of the optical/UV afterglow, while \nfed{} is derived from the optical absorption lines produced by metals at the redshift of the host galaxy ISM. We compare these two quantities to investigate whether $A_V$ is produced in the ISM probed by UV absorption lines. 

Figure \ref{fig: Av_NFedust} shows that \nfed{} steeply increases with $A_V$ in GRB absorbers, at least up to $A_V\sim0.5$. The significance of the \nfed{} correlation with $A_V$ is reported in Table \ref{tab spearman} for the measurements based on both Si and Zn, and also on the individual elements separately. While the correlation seems more solid when including both Zn and Si (2.5$\,\sigma$ significance level), we note that Zn-based measurements alone are not very helpful for the purpose of testing this correlation because of its narrow range of $A_V$. This arises from the observational difficulty of constraining Zn column densities in low-metallicity systems. The scatter of the correlation is quite large, indicating that in some cases there might be some contribution to the $A_V$ that is not related to the bulk of the host-galaxy ISM, as for the clear outlier GRB\,070802 (see below). Overall, the observed trends strongly support the general idea that $A_V$ is mostly produced in the bulk of the host galaxy ISM. A similar conclusion was also reached by \citet{Kruehler11} and \citet{Zafar11}.

GRB\,070802 stands out from the observed trend with the highest $A_V$ measurement. We note that this is one of the few GRBs where a Galactic-like 2175 \AA{} bump was observed imprinted on the optical afterglow spectrum \citep{Eliasdottir09,Zafar12}. The high extinction, well above the general trend of Fig. \ref{fig: Av_NFedust}, might strengthen the hypothesis that in this case the extinction occurs in two different regions with two different types of grains - one of them not representative of the bulk of the host galaxy - as suggested by \citet{Zafar12}. The observation of the 2175 \AA{} bump and the multi-component nature of $A_V$ makes this line of sight stand out from the rest of the GRB sample. One alternative explanation is that the distribution of \nfed{} flattens out, as if some saturation process prevented iron from further accreting on dust grains above a certain threshold.

In general, one obvious candidate for an additional region to contribute to the $A_V$, other than the host ISM, is a dense and dusty environment close to the GRB where no low-ionization lines are produced. However, dust-destruction fronts typically move faster and reach greater distances than the ionization fronts produced by GRBs, making the dust destruction more efficient than ionization \citep{Draine02}. Thus, it seems unlikely to observe regions with dust but no low-ionization gas. Thus, foreground absorbers are more likely to be the source of additional contribution to the $A_V$. However, the positive correlation between \nfed{} and $A_V$ confirms that for most lines of sight the main source of extinction is the GRB host galaxy.

\subsection{Molecular content}

\begin{figure}
   \centering
   \includegraphics[width=90mm,angle=0]{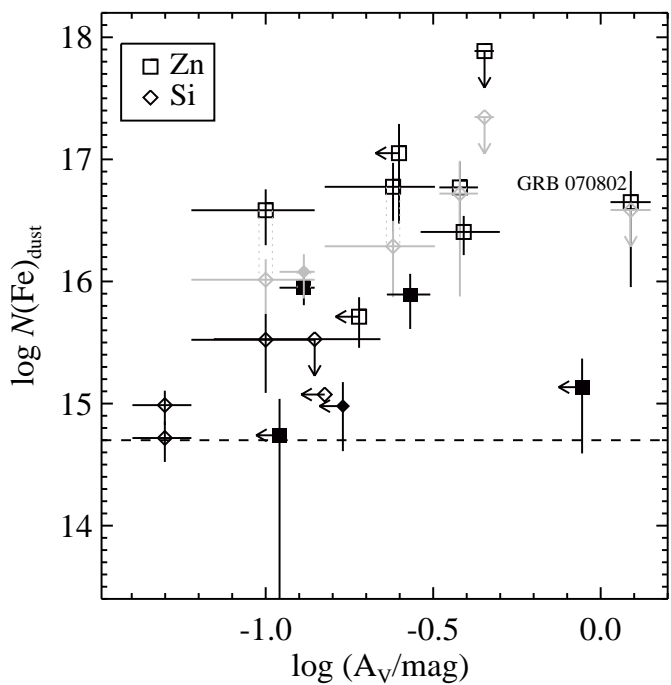}
   \caption{The optical extinction $A_V$ distribution with the column density of Fe in dust. The dashed line marks the column density above which a significant presence of molecules has been observed in QSO-DLAs \citep{Noterdaeme08}. When both are available, the measurements derived from Zn and Si are displayed in black and grey, respectively. The symbols are the same as in Fig. \ref{fig: dtm_distr with Si}.}
              \label{fig: Av_NFedust}
    \end{figure}

Given the GRB association with massive star-forming regions \citep[e.g.,][]{Woosley93,Hjorth12}, we expect to observe the signature of molecular clouds in some optical afterglow spectra. In particular, dusty systems may bear the conditions for molecular formation, since dust grains can be efficient catalysts to aid the formation of molecules, such as H$_2$, on their surfaces \citep[e.g.,][]{Pirronello97}. However, molecular absorption bands are typically weak and notoriously difficult to observe in optical/near-UV spectra of extragalactic sources \citep[e.g.,][]{Noterdaeme08}. In GRB afterglows, CO was detected only once to date in the low-resolution Keck/LRIS spectrum of GRB\,080607 \citep{Prochaska09}, while H$_2$ has been recently observed in a couple of mid-resolution VLT/X-shooter afterglow spectra \citep[][Friis et~al., in preparation]{Kruehler13}. The dashed line in Fig. \ref{fig: Av_NFedust} shows the column density $\log \mbox{\nfed{}}=14.7$ above which molecules have been observed in $\sim40$\% of QSO-DLAs \citep{Noterdaeme08}. \textit{Most GRB-DLAs lie above this column density and thus are promising sites to detect molecules.}

While the presence of molecules is expected in GRB environments, the hard radiation of the bursts are effective in destroying dust and molecules in molecular clouds \citep{Draine02}. However, the currently low detection rate of molecules observed in GRBs may be an observational effect due to the typically low spectral resolution or low S/N of the majority of GRB optical/near-UV spectra. On the other hand, current high-resolution samples might be biased against high metallicity lines of sight (potentially rich in dust and molecules) because only bright afterglows can currently be observed at high spectral resolution \citep{Ledoux09}. A large sample of mid-resolution (e.g., VLT/X-shooter) spectroscopic observations of GRB-DLAs can potentially include a significant number of those systems with the most favorable conditions for molecular formation \citep[i.e., high \hi{} and relatively high metallicity,][]{Noterdaeme08}, with sufficient S/N and spectral resolution to constrain the actual molecular content and the extent of dust destruction.

\section{Conclusions}
\label{conclusions}

We derive the dust-to-metal ratios (\dtm{}) in long-duration GRB- and QSO-DLAs based on the dust-depletion pattern observed from low-ionization metal absorption lines in the optical/UV afterglow spectra. In this way, the dust-to-metal ratio is purely associated with the bulk of the neutral ISM in the GRB- or QSO-DLA, which is not influenced by the GRB, rather than integrated along the total line of sight. As a comparison, we also derive the \dtm{} of GRB absorbers using a different method based on dust-depletion as well, described in \citet{Savaglio01}, with fully consistent results.

The dust-to-metal ratios in GRB-DLAs span $\sim$10--110\% of the Galactic value and show properties similar to the QSO-DLA distribution, with GRBs extending the study of dust-to-metal ratios out to higher column and volume densities. Above metallicities of $Z\sim0.1\,Z_\odot$, these results are partially consistent with the finding of a universally constant Galactic-like dust-to-metal ratio based on $A_V$ \citep{Zafar13}. However, at lower metallicities, down to $Z\sim0.002\,Z_\odot$, we find compelling evidence for low values of \dtm{} for both GRB- and QSO-DLAs. We argue that our \dtm{} estimates for low-metallicity DLAs are more reliable than those based on the dust extinction along the line of sight because low $A_V$ values are more sensitive to the contribution of foreground systems.

The \dtm{} distribution increases with metallicity and metal column density. A similar trend was recently observed by \citet{Brinchmann13} for super-solar metallicity nearby galaxies. We observe a flattening of this trend above $\log N(\mbox{Zn})\sim13$, suggesting that the production/destruction of dust with respect to the metals stabilizes for more evolved galaxies. The low dust fraction of low-metallicity systems may be the result of inefficient dust production, e.g., for grain growth in the low-metallicity regime (where the time scale for grain growth becomes too long), with negligible contribution by SNe and AGB stars.   

A possibility that needs to be further investigated is the presence of a population of low-\dtm{} GRB-DLAs having a smaller amount of dust given their metal content, possibly due to destruction of dust grains in actively star-forming regions.

 At $z\lesssim1.7$, all the \dtm{} estimates for the GRB-DLAs in our sample show high (Galactic-like) values. If confirmed with a larger sample, this may indicate that at lower redshift most GRBs occur in galaxies with $Z\gtrsim0.1Z_\odot$ that are richer in dust with respect to the overall host-galaxies population. This could possibly be due to the overall increase of metallicity with cosmic evolution. The presence of some of these high \dtm{} galaxies is observed out to $z\sim4$, indicating that for these systems a fairly rapid ($\lesssim 1$ Gyr) production mechanism is required, i.e., in cooled SN ejecta as well as in AGB winds and grain growth in a $Z\sim Z_\odot$ metallicity regime. We find no trends of the \dtm{} with the host-galaxy mass, star-formation rate, and specific star-formation rate.

The column density of Fe in dust \nfed{} that we derive from the relative abundances observed in the host-galaxy ISM steeply increases with the dust extinction $A_V$ observed toward GRBs, up to at least $A_V\sim0.5$. This confirms that $A_V$ is mostly produced by the bulk of the ISM neutral gas. At high $A_V$, we observe one outlier of this trend, GRB\,070802, where either an additional contribution for the extinction must be included or the \nfed{} flattens out at high $A_V$. The majority of GRB-DLAs shows high values of \nfed{} \cite[above $\log \mbox{\nfed{}}~14.7$, the minimum value for QSO-DLA with molecular detection][]{Noterdaeme08}, making them promising sites for the detection of molecules.

\begin{acknowledgements}
 We are most grateful to Edward B. Jenkins and Lars Mattsson for insightful comments on the manuscript. We thank Avishay Gal-Yam, Darach Watson and Tayyaba Zafar for useful discussions. ADC acknowledges support by a Grant of Excellence from the Icelandic Research Fund, the Weizmann Institute of Science Dean of Physics Fellowship, and the Koshland Center for Basic Research.
\end{acknowledgements}

\bibliographystyle{aa} 

\bibliography{biblio}

\appendix

\section{Iron as a tracer of the bulk of the dust}
\label{sect iron}
 
Relying on the depletion of iron to trace dust one may wonder whether a bias is introduced, affecting systems that have a significant amount of dust but hardly any iron-rich grains. Here we discuss whether this is a significant issue for our analysis. Iron is typically used as a dust tracer because it is highly depleted \citep[60--90\% of iron is found in dust grains in the Milky Way,][]{Savage96,Pinto13}, while still being easily observable (it is among the most abundant refractory elements in the ISM). After oxygen and carbon, the main dust components in the Milky Way are Mg, Si, and Fe, with about $\sim30$ atoms every $10^6$ H atoms \citep[e.g.,][]{Kimura03,Przybilla08}. Iron traces a variety of dust compounds, such as silicates (both olivine and pyroxenes contain Fe, besides O, Mg, and Si), oxides \citep[Fe embedded in oxides must contribute to silicate mixtures,][]{Voshchinnikov10} and iron-based dust grains. Iron-poor carbonaceous grains contribute to the dust content as well. However, carbonaceous grains are typically more abundant in Galactic-like environments, where iron-rich grains are also well represented. The typical dust extinction of lower-metallicity environments like SMC is well reproduced with silicate grains without a significant contribution from carbonaceous grains \citep{Pei92}. This may apply to GRB host galaxies as well, the majority of which show a SMC- or LMC-type extinction curve \citep{Zafar11,Schady12}. 

Ni and Cr could be used as dust tracers instead of Fe as well, but they are not as diffused and as easily observed as iron. $[$Zn$/$Fe$]$ typically increases with metallicity as a result of dust depletion of Fe in both QSO- and GRB-DLAs \citep[][]{Ledoux02, Savaglio03, Wolfe05, Dessauges06, Prochaska07, Noterdaeme08} and therefore is widely used as a dust indicator. Thus, iron and zinc are the most suitable elements for this analysis and we consider the depletion of iron to be a reliable indicator of the overall amount of dust in GRB host galaxies.

\section{Sulphur}
\label{sect sulphur}
Despite its little depletion onto dust grains, sulphur has been regarded as a ``troublesome element'' because of its unconventional behavior of the depletion in Galactic environments, often leading to negative depletion strength factor $F_*$ \citep{Jenkins09}. In general, the observed depletion of S in the Galaxy is sometimes poorly constrained, for instance for halo environments \citep[\protect{$-0.2\lesssim$ [S/H] $\lesssim0.2$}][]{Savage96}. Besides, S shows an erratic depletion behavior in QSO-DLAs as well where [S/Si] extends down to $\sim-0.3$ \citep[e.g.,][]{Dessauges-Zavadsky06}, either consistent with solar with a very large scatter, or indicating non uniform depletion. If not depletion, either nucleosynthesis (S being an $\alpha$-element) or observational effects are responsible for the observed sulphur behavior. Among the latter, Galactic \suii{} lines are often saturated, while the high-ionization potential of \suii{} (23.3 eV) allows it to survive ionization in \hii{} regions \citep{Jenkins09}. Given these uncertainties, we disregard S as a reliable tracer of dust (non-) depletion for the current study.

\section{Derivation of the Galactic dust-to-metal ratio}
\label{sect dtm_gal}

  \begin{figure*}
   \centering
   \includegraphics[width=120mm,angle=0]{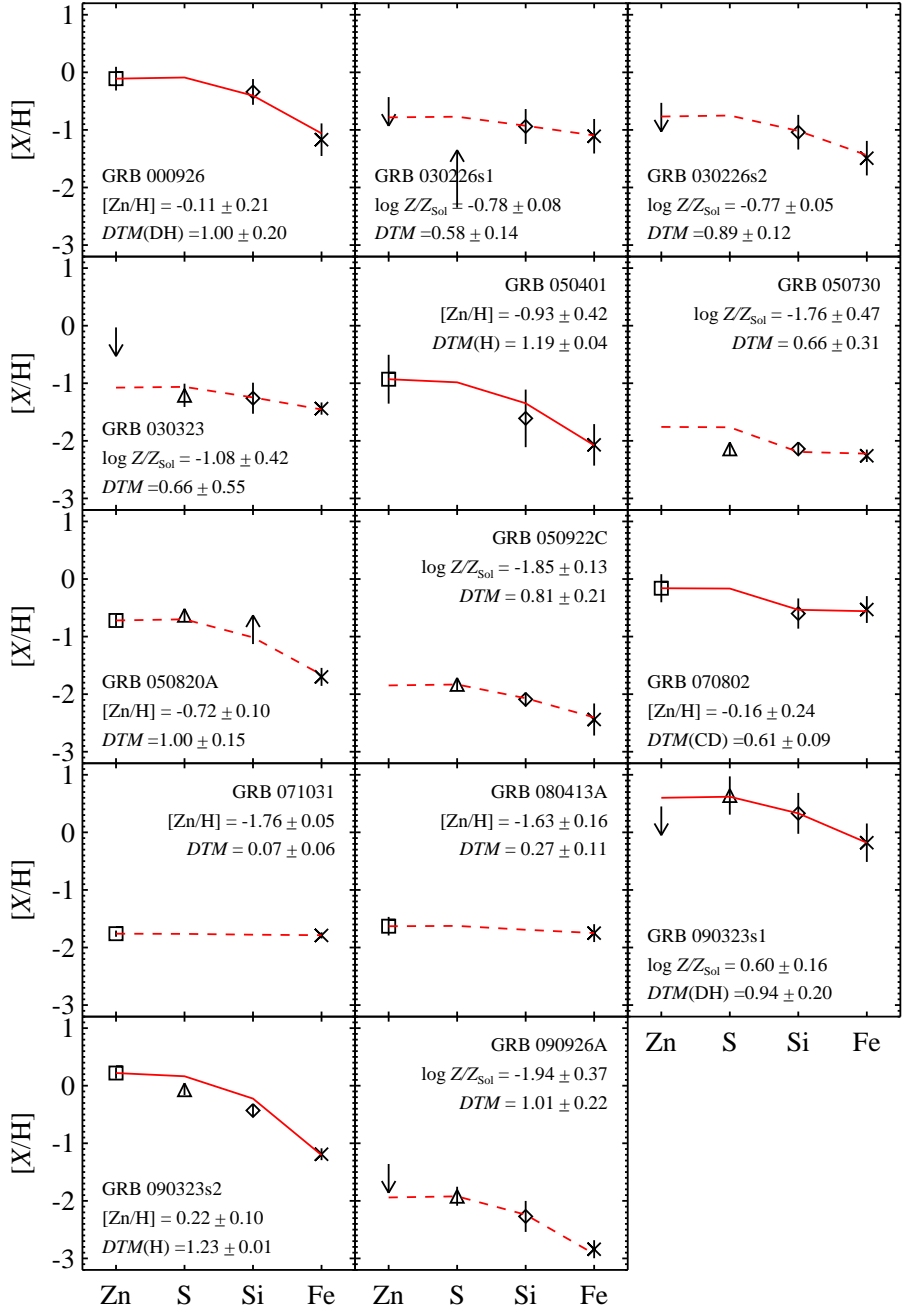}
   \caption{The depletion pattern [$X$/H] observed in GRB absorbers. The solid lines represent the best-fit depletion patterns corresponding to a best-fit dust-to-metal ratio, environment type, and metallicity, labeled on each panel and reported in Table \ref{tab dtm}. [Zn/H] are used as metallicity input for the fit, while $\log Z/Z_\odot$ are the fitted metallicities in case the \hi{} column density, rather than $N$(\znii{}) is constrained. The dashed lines refer to those cases where a single environment type could not be constrained. All upper and lower limits, as well as any constraint on sulphur are not included in the fit, but only displayed for completeness.}
              \label{fig: dtm_sav}
    \end{figure*}

We calculate $dtm$(G) assuming [Fe/Zn]$_G= -0.95$, which is the average value of Galactic disk+halo lines of sight \citep{Savage96} with the $-0.03$ dex correction of \citet{Jenkins09}. The slopes of the Galactic depletion sequences are $d{\rm Fe}_{\rm G} = A_{\rm Fe} = -1.285$ and $d{\rm Zn}_{\rm G} = A_{\rm Zn} = -0.610$ \citep[][where the depletion formalism is expressed in terms of $A$, $B$, and $z$ coefficients]{Jenkins09}. The offsets of the Galactic sequences (i.e., depletion at $F_*= 0$) are ${\rm offset}_{\rm Fe} = B_{\rm Fe} - A_{\rm Fe}\,z_{\rm Fe} = -0.95$ and ${\rm offset}_{\rm Zn}\sim0$, so then 
\begin{subequations}
 \begin{align}       
\delta_{\rm Zn, G} & = d{\rm Zn}_{\rm G} \cdot F_*    \label{eq delta Zn G} \\            
\delta_{\rm Fe, G} & = d{\rm Fe}_{\rm G} \cdot F_* + {\rm offset}_{\rm Fe} \mbox{.}  \label{eq delta Fe G}
\end{align}       
\end{subequations}
Given that $ \delta_{\rm Fe, G} =  \delta_{\rm Zn, G} + {\rm [Fe/Zn]}_{\rm G}$ (see Fig. 1, regardless of the offsets), then Eq. \ref{eq delta Zn G} can be rewritten as
\begin{equation} 
\delta_{\rm Fe, G} -  [{\rm Fe/Zn}]_{\rm G} =     d{\rm Zn}_{\rm G} \cdot F_*\mbox{.}
\label{eq delta Zn G 2} 
\end{equation}
Deriving $F_*$ from Eq. \ref{eq delta Fe G} and substituting in Eq. \label{eq delta Zn G} provides
\begin{equation} 
\delta_{\rm Fe, G} -  [{\rm Fe/Zn}]_{\rm G} =     \frac{d{\rm Zn}_{\rm G}}{d{\rm Fe}_{\rm G}}\,(\delta_{\rm Fe, G} -  {\rm offset}_{\rm Fe} ) \mbox{.} 
 \nonumber
\end{equation}
From $dtm = 1 - 10^{\delta_{Fe}}$ (Eq. \ref{eq dtm}) it finally follows that 
\begin{equation}
dtm({\rm G}) = 1.-10^{    (   [{\rm Fe/Zn}]_{\rm G} \, d{\rm Fe}_{\rm G}   -  {\rm offset}_{\rm Fe} \, d{\rm Zn}_{\rm G}  ) / (dFe_{\rm G} - dZn_{\rm G})  } = 0.89.
\end{equation}

    \begin{figure*}
   \centering
   \includegraphics[width=120mm,angle=0]{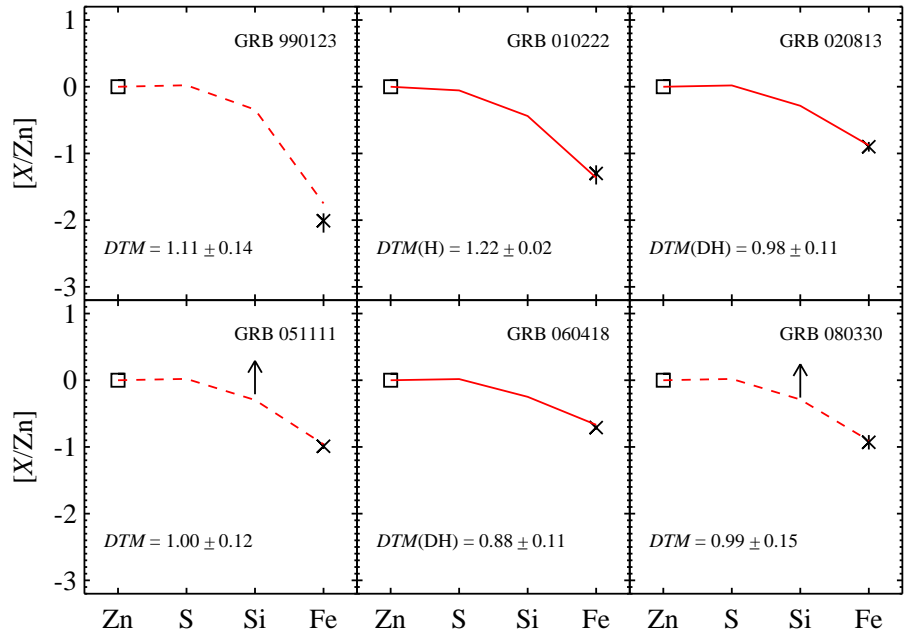}
   \caption{Same as Fig. \ref{fig: dtm_sav}, except for those cases where $N$(\hi{}) is not constrained, so the relative abundances with respect to Zn, [$X$/Zn] are fitted instead of the absolute abundances.}
              \label{fig: dtm_sav_noH}
    \end{figure*}
    
    \begin{figure*}
\centering  
   \includegraphics[width=90mm,angle=0]{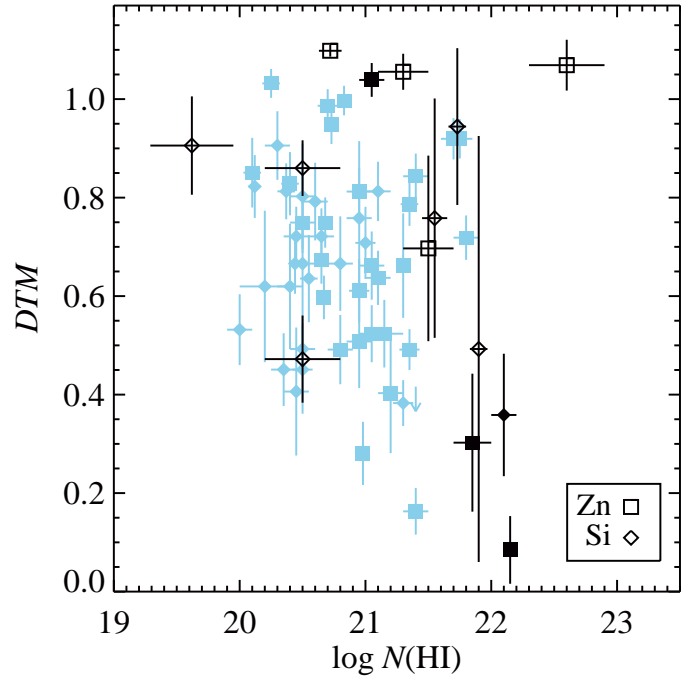}
   \caption{The \dtm{} distribution with the \hi{} column density. The symbols are the same as in Fig. \ref{fig: dtm_distr}.}
              \label{fig: dtm_HI}
    \end{figure*}

\end{document}